\documentclass[a4paper,11pt]{article}

\usepackage{fourier}
\usepackage{color}
 \usepackage{graphicx}
\usepackage{url}
\usepackage[affil-it]{authblk}
\usepackage{amsmath}
\usepackage{wrapfig}
\usepackage{hyperref}
\usepackage[T1]{fontenc}
\usepackage{times}
\usepackage{fancyhdr}
\usepackage{titlesec}
\usepackage{multirow}
\begin{document}
\title{Compact \& Capable: Harnessing Graph Neural Networks and Edge Convolution for Medical Image Classification}

\author{Aryan Singh}
\author{Pepijn Van de Ven}
\author{Ciarán Eising}
\author{Patrick Denny}
\affil{University of Limerick}
\date{}
\maketitle
\thispagestyle{empty}

\begin{abstract}
Graph-based neural network models are gaining traction in the field of representation learning due to their ability to uncover latent topological relationships between entities that are otherwise challenging to identify. These models have been employed across a diverse range of domains, encompassing drug discovery, protein interactions, semantic segmentation, and fluid dynamics research. In this study, we investigate the potential of Graph Neural Networks (GNNs) for medical image classification. We introduce a novel model that combines GNNs and edge convolution, leveraging the interconnectedness of RGB channel feature values to strongly represent connections between crucial graph nodes. Our proposed model not only performs on par with state-of-the-art Deep Neural Networks (DNNs) but does so with 1000 times fewer parameters, resulting in reduced training time and data requirements. We compare our Graph Convolutional Neural Network (GCNN) to pre-trained DNNs for classifying MedMNIST dataset classes, revealing promising prospects for GNNs in medical image analysis. Our results also encourage further exploration of advanced graph-based models such as Graph Attention Networks (GAT) and Graph Auto-Encoders in the medical imaging domain. The proposed model yields more reliable, interpretable, and accurate outcomes for tasks like semantic segmentation and image classification compared to simpler GCNNs. Code available at \href{https://github.com/anonrepo-keeper/GCNN-EC}{AnonRepo}.
\end{abstract}

\section{Introduction}
Medical image classification and segmentation play critical roles in the field of medical imaging. Although there have been considerable advancements in image classification, medical image classification faces unique challenges due to the diverse dataset modalities, such as X-ray, Positron Emission Tomography (PET), Magnetic Resonance Imaging (MRI), Ultrasound (US), and Computed Tomography (CT). Variations within and between modalities, mainly stemming from the inherent differences in imaging technologies, complicate the classification process. Additionally, obtaining labeled training data is costly in the medical domain. Pre-trained DNNs address these issues through transfer learning techniques, yielding impressive results. However, DNNs exhibit limitations, including inductive bias, inefficient capture of spatial and local-level associations, and inconsistent performance across modalities \cite{cnnlim,Zhou_2021}.

\begin{figure*}
 \centering
 \includegraphics[width=0.9\textwidth,height = 6cm]{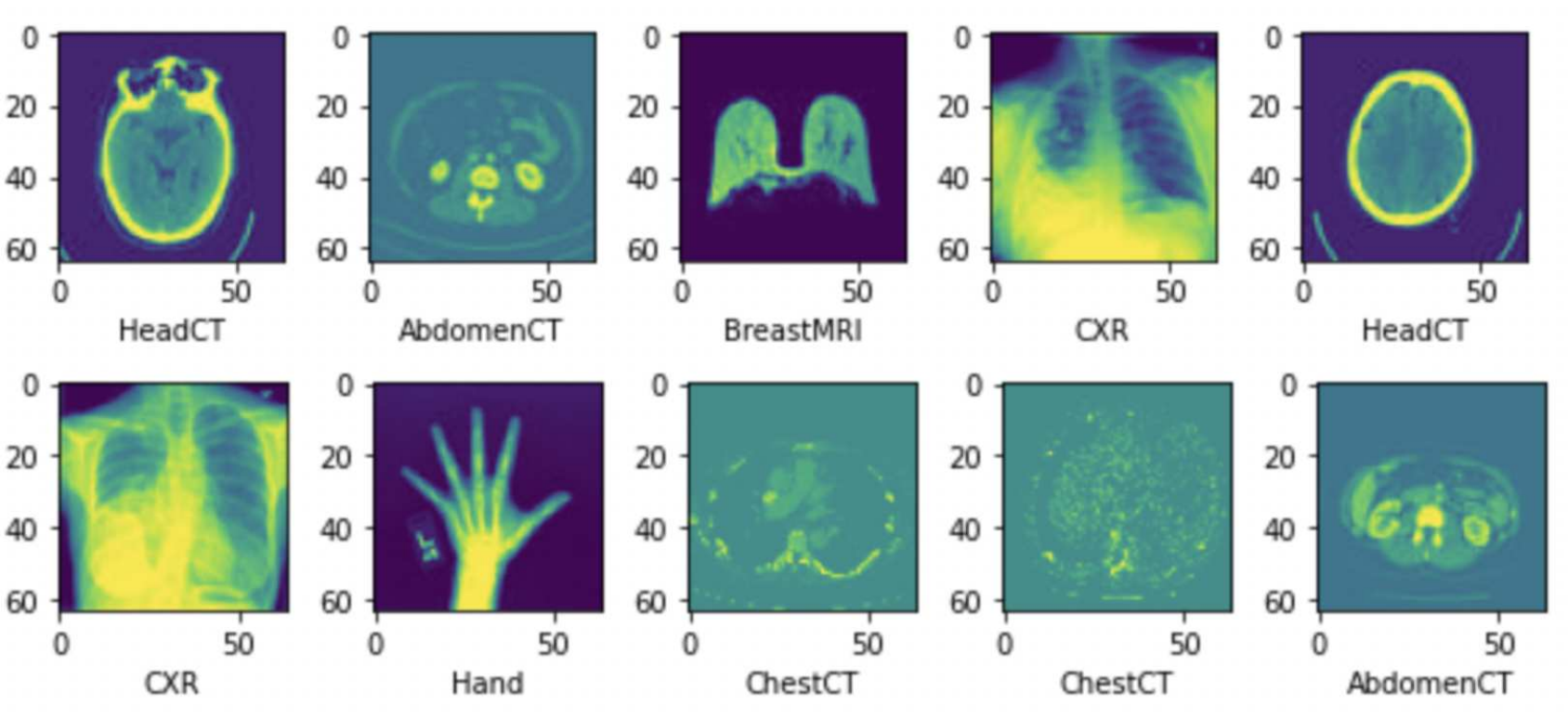}
 \vspace{-4mm}
 \caption{
 Sample MedMNIST data.
 }
 \label{fig:medm}
 \vspace{-0.5cm}
\end{figure*}

GNNs offer a solution to these complexities, handling variations in data with embedded relationships effectively and accommodating heterogeneous graph nodes\cite{hetro1}. The successful application of Knowledge-Based Graph Methods in medical diagnosis supports this notion. We have compared GNN architecture with Convolutional Neural Networks (CNNs) and discussed various types of Graph Convolutional Neural Networks (GCNNs). We propose a GCNN model integrated with Edge Convolution \cite{edcnn}(GCNN-EC) for medical image classification. By performing graph convolution and edge convolution for edge prediction. Edge convolution overcomes the limitations of vanilla GCNN thus improving classification. Our method enhances model performance with reduced training time and data requirements. This research validates graph-based learning's efficiency for medical image data.  In this study, we focus on classifying the MedMNIST dataset \cite{medmnist}, featuring 10 pre-processed datasets from various sources and modalities, with 708,069 images in 12 2D datasets. We narrow our research scope to six categories/classes\ref{fig:medm} containing 58,954 images with dimensions 28x28 as these classes represent diverse modalities, reflecting the compilation of images from various imaging techniques. These classes are AbdomenCT, BreastMRI, CXR, ChestCT, Hand, and HeadCT. The subsets are balanced.

We observe that our simple GCNN-EC outperforms leading state-of-the-art DNNs for specific MedMNIST dataset classes. Proposed model required less training than compared DNNs while using 100 times fewer parameters.  


\vspace{-10pt}

\section{Prior Art}
In this section, we will delve into the technicalities of CNNs, and compare their mechanisms with GCNNs. We will also shed light on three contemporary, state-of-the-art CNN models that have been utilized for the task of medical image classification. Furthermore, this section will introduce the diverse variants of GNNs, providing a comprehensive comparison from a technical standpoint. It will also cover the various applications of these models in medical domains.
\subsection{CNNs}
\begin{wrapfigure}{r}{0.5\linewidth}
 \vspace{-3mm}
 \centering
 \includegraphics[width=0.44\textwidth]{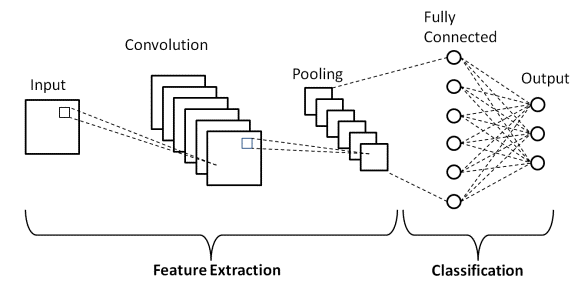}
 \vspace{-3mm}
 \caption{
 CNN architecture.
}
 \label{fig:cnn}
 \vspace{-3mm}
\end{wrapfigure}

CNNs\cite{cnn} owe their name to the convolution operation, which involves overlaying a kernel onto the image grid and sliding it across the grid to extract local information, such as details from neighboring pixels. Technically, the convolution operation involves performing a dot product between the filter's elements and the corresponding elements of the image grid, then storing the result in an output matrix (often termed a feature map or convolved feature). As illustrated in Figure \ref{fig:cnn}, the dot product employed in the convolution process is an aggregation operation. The main objective of this operation is to consolidate image data into a compressed form, making it feasible to extract global-level features from an image.

Thus, convolution as a process systematically extracts spatial hierarchies or patterns, starting from local pixel interactions (low-level features) to more abstract concepts (high-level features) as we progress deeper into the network. Finally, the hierarchical feature extracted from the preceding convolution and pooling operations is compressed into a compact and linear representation. The flattened feature vector derived is used for various tasks such as classification, segmentation, or feature localization.

We have chosen three state-of-the-art DNNs that have demonstrated robust performance in image classification in the medical imaging domain for further discussion in this paper. The effectiveness of our proposed model is assessed in relation to these distinguished DNN models in the later section, thereby offering a comparative study.

\textbf{ResNet} \cite{resnet} is a deep neural network architecture with a varying number of hidden layers, including a large number of convolutional layers to work efficiently by using residual blocks \cite{resnet} that allow the network to effectively learn the residual or the difference between the input and output features. ResNets has been one of the best-performing models on the ImageNet dataset \cite{imagenet} for classification tasks. It has served as a skeleton for several DNNs that continue to use similar skip connection methods for achieving state-of-the-art performance. It has been applied for the classification of medical image data and has proved to produce state-of-the-art results \cite{breast}, the ResNets-based model showed 99.05\% and 98.59\% testing accuracy for binary and multi-class breast cancer recognition.


\par \textbf{DenseNet} \cite{densenet} is one of the densely connected deep-layered neural network architectures that also use residual blocks. They exploit the potential of the deep network by feature reuse, producing more condensed models that are easy to train and highly parameter efficient. Concatenating feature maps learned by different layers increases variation in the input of subsequent layers and improves efficiency. DenseNets uses the Network in Network architecture \cite{nin} which uses multi-layer perceptrons in the filters of convolutional layers to extract more complicated features. DenseNet has increasingly been applied as the backbone model for various medical imaging tasks\cite {Zhou_2021} from image registry to image embedding generation which has further been used for tasks like segmentation and classification.


\par \textbf{EfficientNet} \cite{effnet} makes use of techniques like compound scaling, that enable the efficient scaling of deep neural network architectures to meet specific requirements regarding data or resource limitations. Unlike other deep neural networks, EfficientNet achieves improved model performance without increasing the number of floating-point operations per second (FLOPS), resulting in enhanced efficiency. The method introduces the concept of efficient compound coefficients to uniformly scale the depth, width, and resolution of the network. When scaling a model by a factor of $2^{N}$ in terms of computational resources, EfficientNet scales the network depth by $\alpha^{N}$, network width by $\beta^{N}$, and image size by $\gamma^{N}$. These coefficients, namely $\alpha$, $\beta$, and $\gamma$, are determined through a grid search on the base model. By employing this compound scaling approach, EfficientNet strikes a balance between network accuracy and efficiency. EfficientNet has been proven to produce excellent results for medical image classification \cite{Zhou_2021} even in resource-constrained environments.


\subsection{GNNs}
A GNN is a specialized kind of neural network tailored for handling graph-structured data. It exploits the attributes of nodes and edges to learn representations for nodes, edges, and the overall graph. Its working principle is iterative message passing, where features from neighboring nodes are gathered by each node to update its own feature set. CNNs demonstrate limitations in capturing the associations between features within an image \cite{cnnong}. However, these intricate interconnections can be effectively captured by representing images as graphs and then utilizing GNNs to comprehend these intricate interdependencies. Also, GNNs better capture topological data features compared to CNNs\cite{rep}

The intricate complexities of graph data, ranging from structures as varied as protein sequences and chemical molecules to pixels in medical images serving as nodes, necessitate the transformation of these structures into suitable vector spaces. This transformation is crucial for performing computations and analyses. However, this comes with the daunting task of handling graph isomorphism issues, which involve identifying topological similarities between different graphs. In solving these intricate problems, the significance of GNNs becomes especially apparent, specifically in conjunction with the Weisfeiler-Lehman (WL) Isomorphism algorithm\cite{wlm}.

WL algorithm, a prevalent technique in this context, generates graph embeddings by aggregating colors, or more general features (image features, etc), of proximate nodes, culminating in a histogram-like representation for each graph. The conventional GNN architecture mirrors a neural rendition of the 1-WL algorithm, where '1' represents a single neighborhood hop. In this transformation, discrete colors evolve into continuous feature vectors, and neural network mechanisms are harnessed to aggregate information from the neighborhood of each node.


By virtue of this design, GNNs inherently embody a continuous variant of graph-based message passing similar to the WL algorithm. In this paradigm, details from a node's immediate surroundings are accumulated and relayed to the node, thereby facilitating learning from local graph structures\cite{gnnplus}. This characteristic is at the core of the utility of GNNs in various domains requiring graph-based data analysis. Furthermore, we elaborate in detail on the specific type of GNN employed in this study, namely the GCNN.

\bigbreak
There are two types of GCNNs:

\begin{enumerate}
\item GCNNs based on \textbf{spectral methods} (using convolutions via the convolution theorem \cite{spcnn}). Spectral methods fall into the category of transductive learning, where learning and inference take place on the entire dataset. Spectral CNNs (SCNN) \cite{spcnn} was the first implementation of CNNs on graphs, leveraging the graph Fourier transform \cite{reviewgnn} and defining the convolution kernel in the spectral domain. Examples include the Dynamic Graph Convolutional Network (DGCN), which has been effectively applied to detect relation heat maps in images for pose and gesture detection. HACT-Net \cite{hgcnn} is a further example, which has been applied for the classification of Histopathological images.


\item GCNNs based on \textbf{spatial methods}. These GCNNs fall into the category of inductive learning, where learning and inference can be performed on a test and train dataset. They define convolution as a weighted average function over the neighborhood of the target vertex. For example, GraphSAGE \cite{sage} takes one-hop neighbors as neighborhoods and defines the weighting function as various aggregators over the neighborhood. The spatial GCNN is extremely robust due to its inductive learning which makes spatial GCNNs highly scalable.
\end{enumerate}



\begin{figure*}
 \centering
 \includegraphics[width=0.8\textwidth,height = 4.5cm]{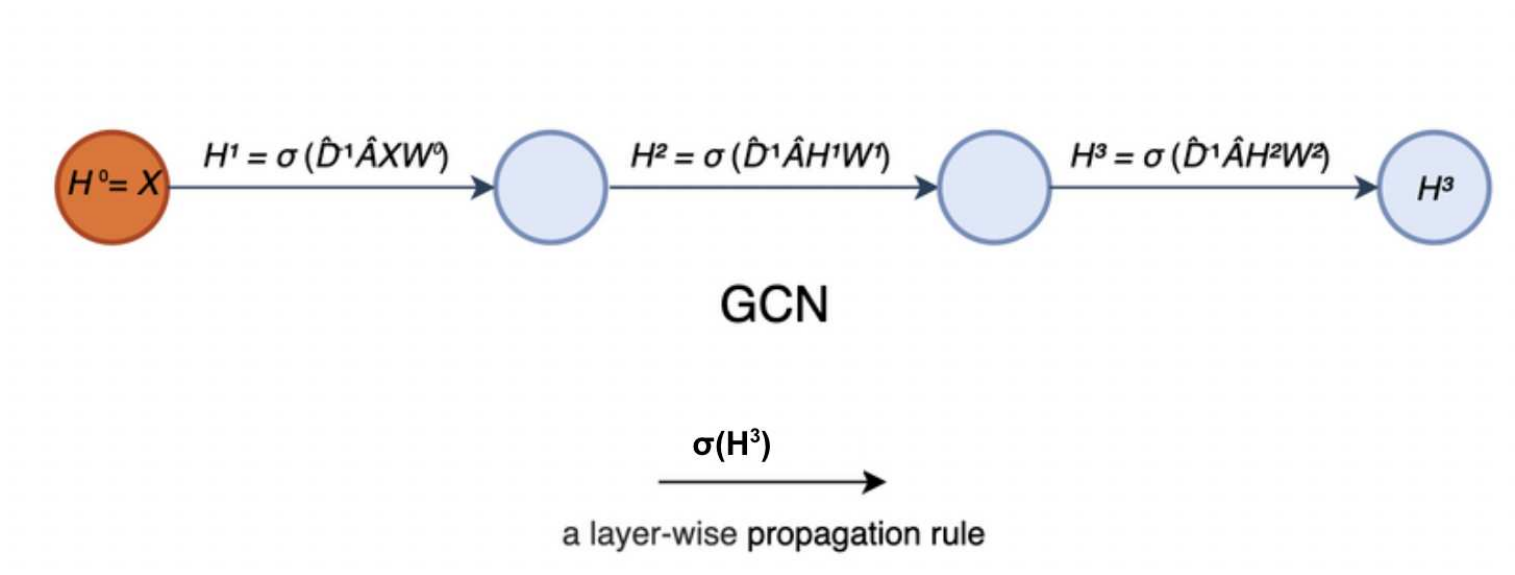}
 \vspace{-4mm}
 \caption{A simple GCN with 3 hidden layers.
 }
 \label{fig:proposed_gnn}
 \vspace{-0.5cm}
\end{figure*}

We use a simple spectral GCNN \(f(X, A)\) that takes input X which is a vector of node features and an adjacency matrix A, along with a layer-wise propagation rule. The matrix ${A}$ is normalized using methods mentioned in \cite{gcnn} as multiplication of ${X}$ and ${A}$ will change the scale of feature vectors, which leads to disproportional learning from neighbors. The equation \ref{eq:eq1} defines the aggregation operation from one layer to another.
\begin{equation}
H^{(l+1)}=\sigma\left(\tilde{D}^{-\frac{1}{2}} \tilde{A} \tilde{D}^{-\frac{1}{2}} H^{(l)} W^{(l)}\right) . \label{eq:eq1}
\end{equation}
\(\tilde{A}\) is the adjacency matrix of the graph \(\mathcal{G}\) , \(\tilde{D}\) is the degree matrix and \(W^{(l)}\) is a trainable weight matrix. The result is passed to an activation function \(\sigma(\cdot)\). The output is then concatenated to get a new hidden state ${H^{(l+1)}}$ for hidden layer ${l+1}$ as shown in Figure \ref{fig:proposed_gnn}.

GCNN has its inherent limitations. They exhibit poor performance when confronted with dynamic graph structures, causing over-fitting to the training set. Furthermore, GCNNs are susceptible to the issue of over-smoothing\cite{gcnlim}, whereby the addition of more convolutional layers results in an indistinguishable final embedding.


In this section, we defined and elaborated on the different types of CNN used in this study along with the types of GNNs and an explanation of spectral GCNN which has been used in the proposed method. In the next section, we explain the proposed method that leverages the power of GNNs, while also overcoming its limitations. 


\section{Our work}
In this work, we present GCNN-EC which resolves identified issues around the limitation of CNNs in capturing the inherent connections between features within an image. The proposed method aims to leverage both local and global inter-pixel relationships by incorporating edge convolution along with graph convolution. 
Our procedure involves three stages: edge convolution, graph convolution, and classification. We merge RGB values into a node feature and compute edge features using a dynamic filter, the system processes the graph representation through multiple convolution layers before flattening the embedding for final classification. We explain these steps in detail in this section.   

\subsection{Edge convolution}
We begin by creating a node feature vector by combining RGB channel values. This is a vital step in transforming the RGB image data into a grid format, where each pixel is a node connected to adjacent pixels. Next, we use a dynamic filter\cite{dync} to learn edge features.  This filter incorporates the node features from the immediate neighbors and also those at a two-hop distance, meaning it considers not just the nearest node but also the nodes that are connected to these nearest nodes. This filter is unique for each input and is learned by the network based on node features and the Euclidean distance between the node feature vectors. This learned filter is stored as a registered buffer in the network and not used during back-propagation. Rather, it is used as an edge feature while being passed through convolution layers. It is through this process that we generate an abstract understanding of the relationships between various components of the image. The graph augmented with node and edge features is further improved by an edge convolution layer \cite{edcn}, which performs convolution operations on the graph using the edge features. The edge convolution layer detects and enhances the edges or boundaries in an image (grid of pixels). It focuses on identifying sharp transitions in intensity/color, which are indicative of object edges. Finally giving a more distinguishable graph representation.

\subsection{Graph convolution}
The graph representation enriched by edge convolution is passed through graph convolution layers, capturing features of the graph by incorporating features of nodes and their connections to finally create a more accurate graph-level embedding. One of the strengths of the graph convolution layer is its ability to incorporate both local and global information. By considering the neighborhood relationships between nodes, it captures local patterns and structures within the graph. Additionally, by aggregating information from neighboring nodes iteratively, it gradually incorporates global information, allowing for a comprehensive understanding of the overall graph. Graph convolution alone suffers from the problem of over-smoothing, This limitation can be overcome by enhancing the graph representation quality by edge convolution to capture meaningful edge information.

\subsection{Classification}
The graph embedding, obtained by flattening the output of the graph convolution layers, is classified using a dense layer. This layer transforms the embedding by matrix multiplication and bias addition thus learning  weighted connections and finally applying non-linear activation. Enabling accurate predictions based on the learned representation of the graph. 

\begin{figure*}
 \centering
 \includegraphics[width=0.9\textwidth,height = 6cm]{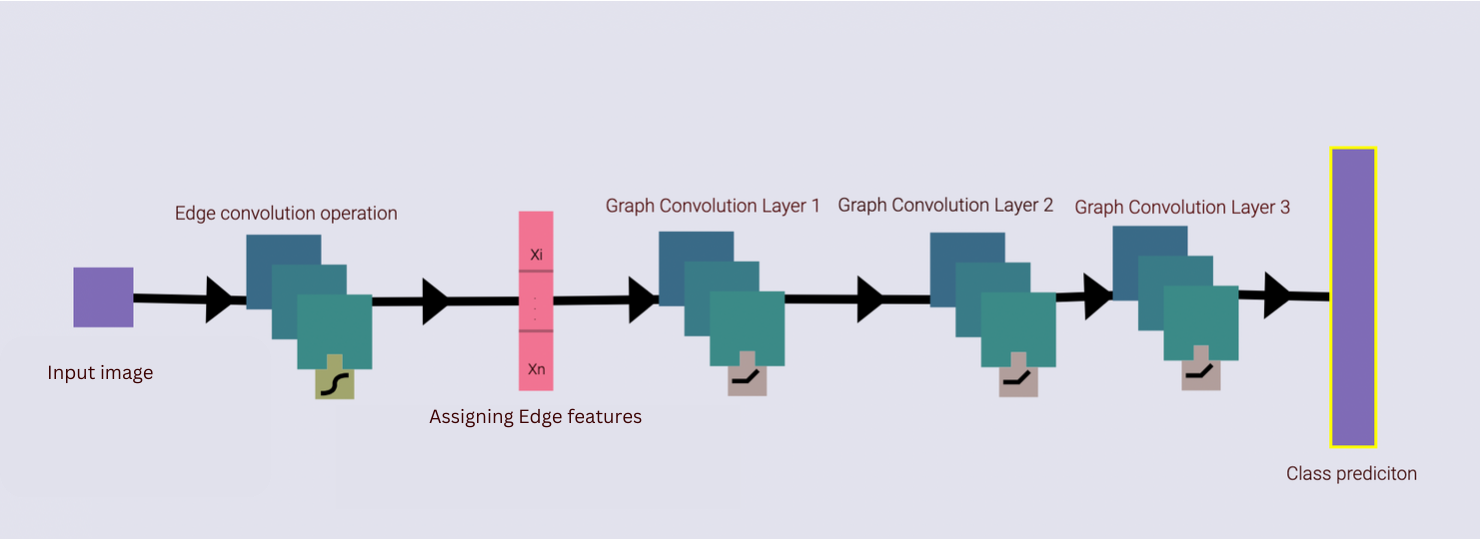}
 \vspace{-4mm}
 \caption{
  GCNN-EC architecture.
 }
 \label{fig:proposed}
 \vspace{-0.5cm}
\end{figure*}

We have employed PyTorch to implement our pipeline, while the Monai framework has been utilized for medical image processing. We generate a graph using MedMNIST images as the input and incorporate the Dynamic Edge Convolution layer\cite{edcn} to perform edge convolution. The resulting learned representation then undergoes 3 graph convolution layers. We fine-tuned the model using Optuna \cite{optuna}, obtaining a learning rate of 0.001 and a weight decay of 0.01. The parameter count in our models ranged from 24,967 to 67,938. We have used the Cross-Entropy loss function with Adam Optimizer and have trained our models for 4 epochs. The batch size used was 64, and the training, testing, and validation splits were 80\%,10\%, and 10\% respectively. GCNN-EC model architecture is shown in Figure\ref{fig:proposed}.



\section{Result}
\begin{figure*}
 \centering
 \includegraphics[width=\textwidth,height = 7cm]{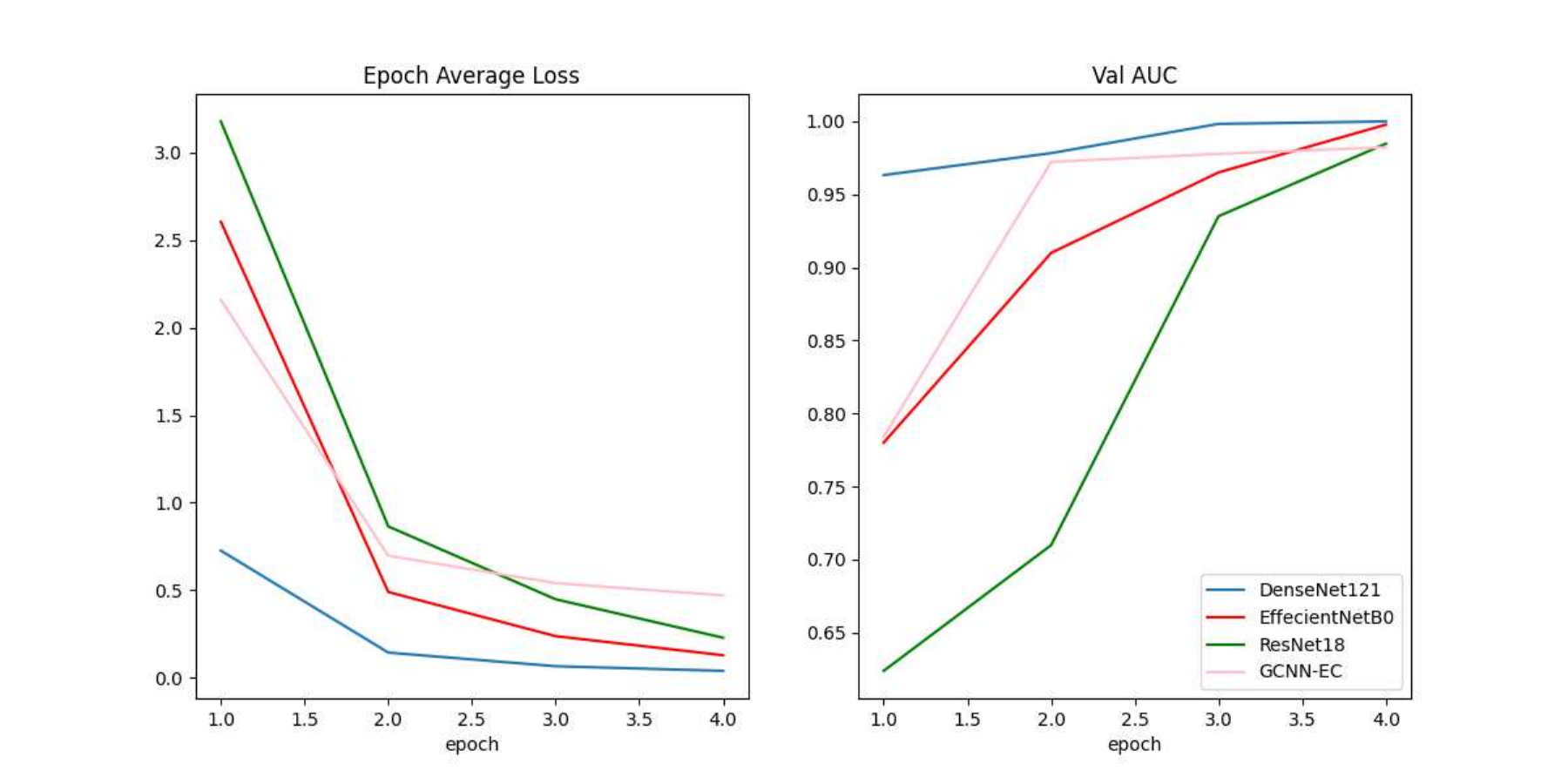}
 \vspace{-4mm}
 \caption{
  Loss and Area under the curve (AUC) over epochs for test dataset.
 }
 \label{fig:auc}
 \vspace{-0.5cm}
\end{figure*}

In this section, we present the results achieved by our model on the 6 classes of the MedMNIST dataset to demonstrate the efficacy of simple GNN when compared with sophisticated DNNs. GCNN-EC model converges to stable loss value within 4 epochs. Our results are presented in Table \ref{tab:tb1} comparing the Area under the Curve (AUC) and Accuracy (ACC) of our model with the DNN models. From the plot in Figure \ref{fig:auc}, it is evident that our method is comparable to DenseNet and outperforms ResNet and EfficientNet, showing it as an effective classifier. The proposed method demonstrates superior performance compared to ResNet18 and EfficientNet-B0 while performing on par with DenseNet121. Notably, the GCNN-EC model utilizes 100 times fewer parameters than the three DNN models considered in this study. Furthermore, our model achieves a remarkable accuracy of 99.13\% on the MNIST dataset (as indicated in the "gcnn-ec-mnist.py" file in the code).

\vspace{-10pt}

\vspace{-10pt}

\begin{table}
    \centering
    \caption{Comparison of DNNs with proposed method for MedMNIST}
    \label{tab:tb1}
    \renewcommand{\arraystretch}{1.4}
    \resizebox{\textwidth}{!}{%
        \begin{tabular}{|l|l|l|l|l|l|l|l|l|l|l|l|l|l|}
        \hline
        \multirow{2}{*}{Methods} & \multicolumn{2}{l|}{AbdomenCT} & \multicolumn{2}{l|}{BreastMRI} & \multicolumn{2}{l|}{CXR} & \multicolumn{2}{l|}{ChestCT} & \multicolumn{2}{l|}{Hand} & \multicolumn{2}{l|}{HeadCT} & \multirow{2}{*}{Parameters} \\ \cline{2-13}
        & AUC & ACC & AUC & ACC & AUC & ACC & AUC & ACC & AUC & ACC & AUC & ACC & \\ \hline
        ResNet18 & 0.800 & 0.839 & 0.897 & 0.899 & 0.832 & 0.842 & 0.901 & 0.940 & 0.915 & 0.921 & 0.733 & 0.762 & 11,689,512 \\ \hline
        EfficientNet-B0 & 0.901 & 0.907 & 0.905 & 0.918 & 0.958 & 0.960 & \textbf{0.913} & \textbf{0.948} & 0.907 & 0.911 & 0.874 & 0.894 & 4,014,658 \\ \hline
        DenseNet121 & \textbf{0.936} & \textbf{0.942} & 0.961 & 0.971 & \textbf{0.972} & \textbf{0.985} & 0.887 & 0.901 & \textbf{0.916} & \textbf{0.925} & \textbf{0.899} & \textbf{0.914} & 7,978,856 \\ \hline
        GCN-EC (ours) & 0.876 & 0.882 & \textbf{0.983} & \textbf{0.985} & 0.957 & 0.965 & 0.748 & 0.813 & 0.886 & 0.905 & 0.869 & 0.874 & \textbf{24,967} \\ \hline
        \end{tabular}%
    }
\end{table}

\vspace{1cm}

\section{Conclusion}
Our model exhibited superior performance when compared to renowned CNNs like ResNet18 and EfficientNet-B0 while achieving comparable results to DenseNet121 on MedMNIST dataset. Notably, our model achieved this with significantly fewer parameters (GCNN-EC: 24,967 vs. ResNet: 11.68M, EfficientNet: 4.01M, DenseNet: 6.95M), highlighting its efficiency and effectiveness in capturing meaningful features. This efficiency suggests the possibility of training our GCNN with significantly fewer data, an important factor in the medical field where properly labeled data is scarce and expensive.


\section*{Acknowledgments}

This publication has emanated from research supported in part by a grant from Science Foundation Ireland under Grant number 18/CRT/6049. For the purpose of Open Access, the author has applied a CC BY public copyright licence to any Author Accepted Manuscript version arising from this submission.

\appendix

\bibliographystyle{apalike}

\bibliography{imvip}

\begin{thebibliography}{}

\bibitem[Akiba et~al., 2019]{optuna}
Akiba, T., Sano, S., Yanase, T., Ohta, T., and Koyama, M. (2019).
\newblock Optuna: A next-generation hyperparameter optimization framework.

\bibitem[Alom et~al., 2018]{breast}
Alom, M.~Z., Yakopcic, C., Taha, T.~M., and Asari, V.~K. (2018).
\newblock Breast cancer classification from histopathological images with
  inception recurrent residual convolutional neural network.

\bibitem[Brabandere et~al., 2016]{dync}
Brabandere, B.~D., Jia, X., Tuytelaars, T., and Gool, L.~V. (2016).
\newblock Dynamic filter networks.

\bibitem[Bronstein et~al., 2017]{rep}
Bronstein, M.~M., Bruna, J., LeCun, Y., Szlam, A., and Vandergheynst, P.
  (2017).
\newblock Geometric deep learning: Going beyond euclidean data.
\newblock {\em IEEE Signal Processing Magazine}, 34(4):18--42.

\bibitem[Bruna et~al., 2013]{spcnn}
Bruna, J., Zaremba, W., Szlam, A., and LeCun, Y. (2013).
\newblock Spectral networks and locally connected networks on graphs.

\bibitem[Defferrard et~al., 2017]{cnnong}
Defferrard, M., Bresson, X., and Vandergheynst, P. (2017).
\newblock Convolutional neural networks on graphs with fast localized spectral
  filtering.

\bibitem[Deng et~al., 2009]{imagenet}
Deng, J., Dong, W., Socher, R., Li, L.-J., Li, K., and Fei-Fei, L. (2009).
\newblock Imagenet: A large-scale hierarchical image database.
\newblock In {\em 2009 IEEE Conference on Computer Vision and Pattern
  Recognition}, pages 248--255.

\bibitem[Dong et~al., 2022]{hgcnn}
Dong, Y., Liu, Q., Du, B., and Zhang, L. (2022).
\newblock Weighted feature fusion of convolutional neural network and graph
  attention network for hyperspectral image classification.
\newblock {\em IEEE Transactions on Image Processing}, 31:1559--1572.

\bibitem[Errica et~al., 2019]{gnnplus}
Errica, F., Podda, M., Bacciu, D., and Micheli, A. (2019).
\newblock A fair comparison of graph neural networks for graph classification.

\bibitem[Hamilton et~al., 2017]{sage}
Hamilton, W.~L., Ying, R., and Leskovec, J. (2017).
\newblock Inductive representation learning on large graphs.

\bibitem[He et~al., 2015]{resnet}
He, K., Zhang, X., Ren, S., and Sun, J. (2015).
\newblock Deep residual learning for image recognition.

\bibitem[Huang et~al., 2016]{densenet}
Huang, G., Liu, Z., van~der Maaten, L., and Weinberger, K.~Q. (2016).
\newblock Densely connected convolutional networks.

\bibitem[Kim et~al., 2023]{hetro1}
Kim, S., Lee, N., Lee, J., Hyun, D., and Park, C. (2023).
\newblock Heterogeneous graph learning for multi-modal medical data analysis.

\bibitem[Kipf and Welling, 2016]{gcnn}
Kipf, T.~N. and Welling, M. (2016).
\newblock Semi-supervised classification with graph convolutional networks.
\newblock {\em CoRR}, abs/1609.02907.

\bibitem[Lecun et~al., 1998]{cnn}
Lecun, Y., Bottou, L., Bengio, Y., and Haffner, P. (1998).
\newblock Gradient-based learning applied to document recognition.
\newblock {\em Proceedings of the IEEE}, 86(11):2278--2324.

\bibitem[Lin et~al., 2014]{nin}
Lin, M., Chen, Q., and Yan, S. (2014).
\newblock Network in network.

\bibitem[Magner et~al., 2020]{gcnlim}
Magner, A., Baranwal, M., and au2, A. O. H.~I. (2020).
\newblock Fundamental limits of deep graph convolutional networks.

\bibitem[Rajpurkar et~al., 2017]{cnnlim}
Rajpurkar, P., Irvin, J., Zhu, K., Yang, B., Mehta, H., Duan, T., Ding, D.,
  Bagul, A., Langlotz, C., Shpanskaya, K., Lungren, M.~P., and Ng, A.~Y.
  (2017).
\newblock Chexnet: Radiologist-level pneumonia detection on chest x-rays with
  deep learning.

\bibitem[Tan and Le, 2019]{effnet}
Tan, M. and Le, Q.~V. (2019).
\newblock Efficientnet: Rethinking model scaling for convolutional neural
  networks.

\bibitem[Wang et~al., 2018]{edcnn}
Wang, Y., Sun, Y., Liu, Z., Sarma, S.~E., Bronstein, M.~M., and Solomon, J.~M.
  (2018).
\newblock Dynamic graph {CNN} for learning on point clouds.
\newblock {\em CoRR}, abs/1801.07829.

\bibitem[Wang et~al., 2019]{edcn}
Wang, Y., Sun, Y., Liu, Z., Sarma, S.~E., Bronstein, M.~M., and Solomon, J.~M.
  (2019).
\newblock Dynamic graph cnn for learning on point clouds.

\bibitem[Weisfeiler and Leman, 1968]{wlm}
Weisfeiler, B.~Y. and Leman, A.~A. (1968).
\newblock The reduction of a graph to canonical form and the algebra which
  appears therein.

\bibitem[Wu et~al., 2021]{reviewgnn}
Wu, Z., Pan, S., Chen, F., Long, G., Zhang, C., and Yu, P.~S. (2021).
\newblock A comprehensive survey on graph neural networks.
\newblock {\em {IEEE} Transactions on Neural Networks and Learning Systems},
  32(1):4--24.

\bibitem[Yang et~al., 2021]{medmnist}
Yang, J., Shi, R., and Ni, B. (2021).
\newblock {MedMNIST} classification decathlon: A lightweight {AutoML} benchmark
  for medical image analysis.
\newblock In {\em 2021 {IEEE} 18th International Symposium on Biomedical
  Imaging ({ISBI})}. {IEEE}.

\bibitem[Zhou et~al., 2021]{Zhou_2021}
Zhou, S.~K., Greenspan, H., Davatzikos, C., Duncan, J.~S., Ginneken, B.~V.,
  Madabhushi, A., Prince, J.~L., Rueckert, D., and Summers, R.~M. (2021).
\newblock A review of deep learning in medical imaging: Imaging traits,
  technology trends, case studies with progress highlights, and future
  promises.
\newblock {\em Proceedings of the {IEEE}}.

\end{thebibliography}

\end{document}